\documentstyle[aps,epsfig,multicol]{revtex}
\begin{document}
\title{Superfluid response in monolayer high-$T_c$ cuprates}

\author{C. Panagopoulos$^1$, T. Xiang$^2$, W. Anakool$^1$, J.R. Cooper$^1$,
Y.S. Wang$^3$ and C.W. Chu$^3$}

\address{$^1$ Cavendish Laboratory and IRC in Superconductivity, University of
Cambridge, Cambridge CB3 0HE, United Kingdom}

\address{$^2$ Institute of Theoretical Physics and Interdisciplinary Center of
Theoretical Studies, Chinese Academy of Sciences, P.O. Box 2735, Beijing 100080,
China}

\address{$^3$ Department of Physics and Texas Center for Superconductivity,
University of Houston, Houston, Texas 77204-5932}

\date{\today}
\maketitle

\begin{abstract}
We have studied the doping dependence of the in-plane and
out-of-plane superfluid density, $\rho^{s}(0)$, of two monolayer
high-$T_{c}$ superconductors, HgBa$_2$CuO$_{4+\delta}$ and
La$_{2-x}$Sr$_x$CuO$_4$, using the low frequency
$ac$-susceptibility and the muon spin relaxation techniques. For
both superconductors, $\rho^{s}(0)$ increases rapidly with doping
in the under- and optimally doped regime and becomes nearly doping
independent above a critical doping, $p_{c}\sim 0.20$.
\end{abstract}


\begin{multicols}{2}

Measurements of the magnetic penetration depth have been important
in probing the order parameter and in testing theories of
high-$T_{c}$ superconductors (HTS) \cite{WNH,Porch,Xiang}. In hole
doped HTS, the low temperature dependence of the in-plane
penetration depth $\lambda_{ab}(T)$ is linear and doping
independent, indicating the presence of nodes in the
superconducting energy gap \cite{WNH,CP1}. The $c$-axis
penetration depth $\lambda _{c}$ is a key parameter for some
theories describing the mechanism of high temperature
superconductivity \cite{JMW,PWA,SC,TX,CP2,KAM,AAT,JRK,DB}. It is
sensitive to the electromagnetic anisotropy of the system and has
been used to test the pairing symmetry and properties of
interlayer dynamics \cite{Xiang,TX,CP2}.

In a recent study of the spin and charge response of HTS, it was
found that both the superfluid density $\rho^{s}(0)$ $\sim$
$\lambda^{-2}(0)$ and the muon spin relaxation, $\mu $SR, rate
show dramatic changes at a critical doping $p_{c}\sim 0.20$,
slightly above optimal doping, in pure and Zn-doped
La$_{2-x}$Sr$_{x}$CuO$_{4}$ (La-214) and
Bi$_{2.1}$Sr$_{1.9}$Ca$_{1-x}$Y$_{x}$Cu$_{2}$O$_{8+y}$ (Bi-2212)
at zero temperature \cite{CP3}. The sharp changes in the
superfluid density with the disappearance of a spin glass phase
transition near $p_c$ suggested a change in symmetry of the ground
state. The existence of such a special doping has been
demonstrated in many other physical quantities \cite{JLT1} and the
$\rho^{s}(0)$ and $\mu $SR data could be linked to the presence of
a quantum phase transition at $p_c$, that is in turn related to
the opening of the normal state pseudogap.

To elucidate further the changes in the ground state across the
phase diagram of HTS we have studied the doping dependence of the
zero-temperature in-plane and out-of-plane superfluid responses,
$\rho _{ab}^{s}$ and $\rho _{c}^{s}$, for two monolayer HTS
materials: HgBa$_{2}$CuO$_{4+\delta}$ (Hg-1201) and La-214. This
study allows us to determine both the in- and out-of-plane
responses as a function of doping and to perform a direct
comparison between two simple HTS with different degrees of
disorder. We find that in both systems, the superfluid density is
strongly doping dependent below $p_{c}$ and shows abrupt changes
around $p_{c}$. For Hg-1201 the effect is sharper and there is
actually a peak in the superfluid density at $p_c$.

The Hg-1201 samples were prepared in Houston using a method
similar to that described in ref. \cite{QX}. Their doping level
can be continuously varied from very under- to heavily over-doped
regime by adding or removing oxygen. Unlike La-214, where the
doping is varied by Sr substitution for La, which may cause
pronounced disorder effects, the variation of oxygen concentration
in Hg-1201 is known to induce little lattice disorder \cite{JPA}.
The Hg-1201 samples were characterised using magnetisation and
thermoelectric power measurements. The doping level $p$ was
determined by both thermopower \cite{Obertelli} and the universal
relation $T_c$=$T_{c,max}$[1-82.6($p$-0.16)$^2$] \cite{Presland}.
The La-124 samples were synthesised in Cambridge using solid-state
reaction followed by oxygenation. Effort was taken to ensure high
purity and homogeneity. All La-214 powders were dried, reacted,
ground, milled, re-pressed and re-sintered at least four times.
The phase purity was verified by powder x-ray diffraction as well
as extensive transport and thermodynamic measurements. No signal
of impurities or inhomogeneity was captured in micro-analytical
spectroscopic studies \cite{Lampakis}. The $T_{c}$ values as well
as lattice parameters of these samples were in good agreement with
published data. In La-214 $p$ is taken as equal to the Sr
concentration. The heat capacity anomalies and $ac$
-susceptibility transitions are sharp.

We have measured the magnetic penetration depth $\lambda $ using
the low-field $ac$-susceptibility technique for grain-aligned
powders \cite{CP2,CP4}. The superfluid density is inversely
proportional to the square of the in-plane penetration depth. To
determine the in-plane and $c$-axis penetration depths separately,
the grains were magnetically aligned in a static field of 12T at
room temperature. X-ray powder diffraction scans \cite{Chrosch}
for both La-214 and Hg-1201 samples showed that more than $90\%$
of the grains had their CuO$_{2}$ planes aligned. The
$ac$-susceptibility measurements were performed down to 1.2K with
a home-made susceptometer using miniature coils with $H_{ac}$=1-3G
rms at $f$=333Hz. The absence of weak links among grains was
confirmed by the linear response of the signal with $H_{ac}$ from
0.3 to 3G rms and $f$ from 33 to 333Hz. We also used a commercial
susceptometer to confirm some of our findings. Taking the grains
to be approximately spherical, as indicated by scanning electron
microscopy, the data were analysed using London's model
\cite{Porch,DS}. The $ac$-susceptibility data were also confirmed
by standard transverse field $\mu$SR experiments performed on
unaligned powders at 400G \cite{CP5}.

Figure 1 shows the data for (a) $T_{c}$, $\lambda_{ab}^{-2}(0)$
and (b) $\lambda _{c}^{-2}(0)$ for La-214. The $T_{c}$ and
$\lambda_{ab}^{-2}(0)$ data were published in an earlier paper and
are included here for comparison \cite{CP3}. $\rho _{ab}^{s}(0)$
is nearly doping independent in the overdoped regime, but drops
fast below $p=0.19$. This suppression of the superfluid density
for $p<0.19$ was previously discussed in terms of a competition
between quasi-static magnetic correlations and superconductivity
\cite{CP3}. It has also been linked to the strong reduction in
entropy as well as condensation energy associated with the opening
of the normal state pseudogap \cite{JLT1,CP5}.

$\rho _{c}^{s}(0)$ shows similar behaviour as its in-plane
counterpart. However, in contrast to the nearly linear doping
dependence of $\rho _{ab}^{s}(0)$ on $p$, $\rho _{c}^{s}(0)$ shows
a stronger doping dependence below $p_{c}$ corresponding to $1/
\lambda_c^{2}$ $\propto$ $p^n$ with $n \sim 2.7$. This difference
in the doping dependence between $\rho_{ab}^{s}(0)$ and $\rho
_{c}^{s}(0)$ is probably associated with the unconventional
interlayer coupling of electrons in high-$T_{c}$ oxides, and is
worthy of further theoretical and experimental investigation.

Figure 2(a) shows the doping dependence of $T_{c}$ and
$\rho_{ab}^{s}(0)$ for Hg-1201. Similar to La-214, $\rho
_{ab}^{s}(0)$ is relatively doping independent in the overdoped
regime and shows a sharp drop below 0.19. A similar $p$ dependence
of $\rho_{ab}^{s}$ has been found for Bi-2212 \cite{WA}, and
recently also reported for
Y$_{0.8}$Ca$_{0.2}$Ba$_{2}$Cu$_{3}$O$_{7-\delta }$ (Ca:Y-123) and
$Tl_{0.5-y}Pb_{0.5+y}Sr_{2}Ca_{1-x}Y_{x}Cu_{2}O_{7}$ (Pb:Tl-2212)
\cite{CB}. The maximum of $\rho _{ab}^{s}(0)$ is located at
$p_{c}$ for all high-$T_{c}$ compounds. It suggests that the
observed doping dependence of $\rho _{ab}^{s}(0)$ below $p_{c}$ is
common to all HTS compounds and is not due to a structural
transition or inhomogeneity. The relatively doping independent
$\rho_{ab}^{s}(0)$ for $p>p_{c}$ in La-214 and Hg-1201 is in
agreement to Bi-2212 \cite{WA} but seems to differ from the data
for $Tl_{2}Ba_{2}CuO_{6+\delta }$, Ca:Y-123 and Pb:Tl-2212
\cite{CB,YJU}. The mechanism causing this difference is unknown
and is certainly worth further investigation. Nevertheless, it is
clear that the maximum of $\rho _{ab}^{s}(0)$ is located at
$p_{c}$ for all high-$T_c$ cuprates.

Figure 2(b) shows the doping dependence of the $c$-axis superfluid
density for Hg-1201. A sharp change from large to low superfluid
response is also observed around $p_{c}$. This is the sharpest
change in $\rho_{c}^{s}(0)$ ever being reported and together with
the observed peak at $p_c$ could be related to its tetragonal
crystal structure and the fact that Hg-1201 is more ordered than
La-214. It is worth noting that a significantly weaker glassy
response has been observed in Hg-1201 \cite{CP6}. We may speculate
that this observation suggests that the sharper changes near $p_c$
may be linked to a quantum critical point for which disorder
causes smoothing of the doping dependence of various physical
properties and associated phase transitions.

The interlayer distance between the CuO$_{2}$ planes may be a key
parameter for optimal $T_{c}$. This has been emphasised by Uemura
recently \cite{YJU}. Indeed, for the same in-plane superfluid
density, $T_{c}$ is higher if the interlayer distance is shorter.
Therefore, the interlayer coupling seems to be essential for
obtaining higher $T_{c}$. The observed variation of $T_{c}$ cannot
be explained by the simple Kosterlitz-Thouless transition where
$T_{c}$ is solely determined by the 2D superfluid density. The
similar doping dependence of $\rho _{c}^{s}(0) $ to $\rho
_{ab}^{s}(0)$ observed here supports this view and indicates the
fundamental role of the $c$-axis electrodynamics to the overall
superconducting condensation. As a matter of fact, $\lambda
_{c}(0)$ for both monolayer cuprates studied here is small above
$p_{c}$ and agrees, for example, with the interlayer tunneling
model of Anderson and Chakravarty \cite{JMW,PWA,SC}. Large
superfluid response above $p_c$ seems to occur in connection with
a crossover from two-dimensional to three-dimensional transport,
as suggested by the doping dependence of the anisotropy in
$\lambda$ (Fig. 3) and the associated behaviour of the anisotropy
of the normal state resistivity \cite{JRC,Takenaka}.

In summary, for the two monolayer high-$T_{c}$ cuprates, La-214
and Hg-1201, both the in-plane and $c$-axis superfluid response
remain relatively constant above $p_{c}$ but drop rapidly below
$p_{c}$. We have found a peak in $\rho^{s}(0)$ at $p_c$ for
Hg-1201 indicating the strongest superconductivity at the point
where the spin glass phase transition (the glass transition
temperature $T_g$ vesus $p$ curve) vanishes and the normal state
gap extrapolates to zero \cite{CP3}. The rapid change and peak may
be due to a change in the superconducting ground state.
Furthermore, we have observed that the doping dependence of
$\rho_c^{s}(0)$ in La-214 follows a power law of approximately
2.7.

C.P. thanks S. Chakravarty and J. W. Loram for enlightening
discussions, D.N. Basov for an earlier collaboration and
discussions on the subject and The Royal Society for financial
support. T.X. acknowledges support from the National Natural
Science Foundation of China.

\begin{figure}[h]
\begin{center}
\caption{(a) Doping dependence of the superconducting transition
temperature $T_c$ and inverse square of the zero temperature
in-plane penetration depth for La$_{2-x}$Sr$_x$CuO$_4$ (La-214)
measured by the $ac$-susceptibility technique. (b) Doping
dependence of the inverse square of the zero temperature
out-of-plane penetration depth. }
\end{center}
\end{figure}

\begin{figure}[h]
\begin{center}
\caption{(a) Doping dependence of the critical temperature $T_c$
and inverse square of the zero temperature in-plane penetration
depth for HgBa$_{2}$CuO$_{4+\delta}$ (Hg-1201). (b) Doping
dependence of the inverse square of the zero temperature
out-of-plane penetration depth. }
\end{center}
\end{figure}

\begin{figure}[h]
\begin{center}
\caption{Doping dependence of the anisotropic ratio
$\lambda_c(0)$/$\lambda_{ab}(0)$ for La$_{2-x}$Sr$_x$CuO$_4$
(La-214) and HgBa$_{2}$CuO$_{4+\delta}$ (Hg-1201).}
\end{center}
\end{figure}

\end{multicols}


\begin{references}

\bibitem{WNH} W.N. Hardy, D.A. Bonn, D.C. Morgan, R. Liang, and
K.Zhang, Phys. Rev. Lett. {\bf 70}, 3999 (1993).

\bibitem{Porch} A. Porch, J.R. Cooper, D.N. Zheng, J.R. Waldram,
A.M. Campbell, and P.A. Freeman, Physica C {\bf 214}, 350 (1993).

\bibitem{Xiang} T. Xiang, C. Panagopoulos, and J.R. Cooper, Int. J.
Mod. Phys. B {\bf 12}, 1007 (1998).

\bibitem{CP1} C. Panagopoulos and T. Xiang, Phys. Rev. Lett. {\bf 81},
2336 (1998).

\bibitem{JMW} J.M. Wheatley, T. Hsu, and P.W. Anderson, Nature (London)
{\bf 33}, 121 (1988).

\bibitem{PWA} P.W. Anderson, Science {\bf 256}, 1526 (1992).

\bibitem{SC} S. Chakravarty, A. Sudbo, P.W. Anderson, and S. Strong,
Science {\bf 261}, 337 (1993).

\bibitem{TX}  T. Xiang, and J.M. Wheatley, Phys. Rev. Lett. {\bf 77}, 4632 (1996).

\bibitem{CP2} C.Panagopoulos, J.R. Cooper, T. Xiang, G.B. Peacock, I. Gameson,
and P.P. Edwards, Phys. Rev. Lett. {\bf 79}, 2320 (1997).

\bibitem{KAM} K.A. Moler, J.R. Kirtley, D.G. Hinks, T.W. Li, and M. Xu,
Science {\bf 279}, 1193 (1998).

\bibitem{AAT} A.A. Tsetkov $et al$., Nature (London) {\bf 395}, 360 (1998).

\bibitem{JRK} J.R. Kirtley $et al$., Phys. Rev. Lett. {\bf  81}, 2140 (1998).

\bibitem{DB} S.V. Dordevic, E.J. Singley, D.N. Basov, S. Komiya, Y. Ando,
E. Bucher, C.C. Homes, and M. Strongin, Phys. Rev. B {\bf 65}, 134511 (2002).

\bibitem{CP3} C. Panagopoulos, J.L. Tallon, B.D. Rainford, T. Xiang, J.R. Cooper,
and C.A. Scott, Phys. Rev. B {\bf 66}, 064501 (2002).

\bibitem{JLT1} J.L. Tallon and J.W. Loram, Physica C {\bf 349}, 53 (2001).

\bibitem{QX} Q. Xiong $et al$., Phys. Rev. B {\bf 50}, 10346 (1994).

\bibitem{JPA} J.P. Attfield, A.L. Kharlanov, and J.A. McAllister, Nature (London)
{\bf 394}, 157 (1998).

\bibitem{Obertelli} S.D. Obertelli, J.R. Cooper, and J.L. Tallon, Phys. Rev. B
{\bf 46}, 14928 (1992).

\bibitem{Presland} M.R. Presland, J.L. Tallon, R.G. Buckley, R.S. Liu, and N.E.
Flower, Physica C {\bf 176}, 95 (1991).

\bibitem{Lampakis}  D. Lampakis, D. Palles, E. Liarokapis, C. Panagopoulos,
J.R. Cooper, H. Ehrenberg, and T. Hartmann, Phys. Rev. B {\bf 62},
8811 (2000).

\bibitem{CP4} C. Panagopoulos, J.R. Cooper, G.B. Peacock, I. Gameson, P.P. Edwards,
W. Schmidbauer, and J.W. Hodby, Phys. Rev. B {\bf 53}, R2999 (1996).

\bibitem{Chrosch} J. Chrosch, C. Panagopoulos, N. Athanassopoulou, J.R. Cooper,
and E.K.H. Salje, Physica C {\bf 265}, 233 (1996).

\bibitem{DS} D. Shoenberg, $Superconductivity$ (Cambridge University Press,
Cambridge, 1954), p. 164.

\bibitem{CP5} C. Panagopoulos, B.D. Rainford, J.R. Cooper, W. Lo, J.L. Tallon,
J.W. Loram, J. Betouras, Y.S. Wang, and C.W. Chu, Phys. Rev. B
{\bf 60}, 14617 (1999).

\bibitem{WA} W. Anakool $et al$., $unpublished$.

\bibitem{CB} C. Bernhard, J.L. Tallon, T. Blasius, A. Golnik, and C. Niedermayer,
Phys. Rev. Lett. {\bf 86}, 1614  (2001).

\bibitem{YJU} Y.J. Uemura $et$ $al.$, Phys. Rev. Lett. {\bf 62}, 2317 (1989).

\bibitem{CP6} C. Panagopoulos, $unpublished$.

\bibitem{YJU} Y.J. Uemura, Solid State Comm. {\bf 126}, 23 (2003).

\bibitem{JRC} J.R. Cooper, L. Forro, G. Collin, and J.Y Henry, Solid State Comm.
{\bf 75}, 737 (1990).

\bibitem{Takenaka} K. Takenaka, K. Mizuhashi, H. Takagi, and S. Uchida, Phys. Rev. B
{\bf 50}, 6534 (1994).


\end{references}
\end{document}